\documentclass[aps,prd,twocolumn,groupedaddress,showpacs,showkeys]{revtex4}

\usepackage{graphicx}
\usepackage{graphicx}
\usepackage{amssymb}
\usepackage{amsmath}
\usepackage{color}
\usepackage{float}
\newcommand{\be}{\begin{equation}}
\newcommand{\ee}{\end{equation}}
 
 \definecolor{BrickRed}{cmyk}{0,0.89,0.94,0.28}
\definecolor{MidnightBlue}{cmyk}{0.98,0.13,0,0.43}
\definecolor{DarkGreen}{rgb}{0,0.7,0.1}

\begin{document}

\title{Interplay of curvature and temperature in the Casimir-Polder interaction}

\date{\today}

\author{Giuseppe Bimonte$^{1,2}$ and Thorsten Emig$^{3,4}$}  

\affiliation{${ }^{1}$Dipartimento di Scienze Fisiche, Universit\`{a} di
Napoli Federico II, Complesso Universitario
di Monte S. Angelo,  Via Cintia, I-80126 Napoli, Italy\\
${ }^{2}$ INFN Sezione di Napoli, I-80126 Napoli, Italy\\
${ }^3$ Massachusetts Institute of Technology, MultiScale Materials Science for Energy and Environment, Joint MIT-CNRS Laboratory (UMI 3466), Cambridge, Massachusetts 02139, USA\\
${ }^{4}$ Massachusetts Institute of
Technology, Department of Physics, Cambridge, Massachusetts 02139, USA}

\begin{abstract}
  We study the Casimir-Polder interaction at finite temperatures between a polarizable small, anisotropic particle and a non-planar surface using a derivative expansion. We obtain the leading and the next-to-leading curvature corrections to the interaction for low and high temperatures. Explicit results are provided for the retarded limit in the presence of a perfectly conducting surface.
\end{abstract}


\maketitle

\section{Introduction}

Dispersion forces, i.e., long-range forces between polarizable bodies originating from quantum and  thermal fluctuations of the electromagnetic (em) field, still constitute a very active field of research thanks to modern experimental techniques, which allow to observe these forces with unprecedented precision. Despite  the similarity of the underlying physical mechanisms, dispersion forces are customarily denoted by different names (van der Waals, London, Casimir-Polder), depending on the size of the involved bodies and the separation scales.
For reviews on dispersion forces, see \cite{parse,rev1,rev2}. In what follows we shall use the abbreviations PP and PS to denote in general the dispersion interaction  of a small anisotropic  particle (including atoms and molecules), with another particle or with a surface, respectively, while the abbreviation SS shall denote the interaction among two surfaces. 

Among the three above manifestations of dispersion forces, PP forces are by far the simplest, because at separations that are large compared to their sizes, small dielectric particles can be modelled as simple dipoles. The situation gets much more complicated however when one or two macroscopic surfaces are involved, as it happens in the PS and in the SS cases respectivley, for then the characteristic many-body nature of dispersion forces entailed by their long-range character  makes it very difficult to compute the  interaction as soon as one (or two) non-planar surfaces are involved. Recent advances in Micro and Nano-Electro-Mechanical Systems prompted a strong interest in the study of  geometry effects on dispersion forces. In the PS case, which is the specific object of this paper, 
geometry effects are important for devices used to trap atoms or molecules close a surface. For a review of recent experiments involving the CP interaction of atoms with microstructured surfaces, see \cite{exp1,exp2,exp3,exp4}. 

Due to the difficulty of the problem, theoretical investigations of the PS interaction  for non planar surfaces have been carried  out only for a few specific geometries.
The example of a uniaxially  corrugated surface was studied numerically in Ref.~\cite{babette} within a toy scalar field theory, while a rectangular dielectric gratings were considered in Ref.~\cite{dalvit}. In Refs.~\cite{galina,pablo} analytical results were obtained for the case of a perfectly conducting cylinder.   A perturbative approach is presented in~\cite{messina},  where surfaces with smooth corrugations of any shape, but with small amplitude, were studied. The validity of the latter is  restricted to particle-surface separations that are much larger than the corrugation amplitude. An alternative approach that  becomes exact in the opposite limit of small particle-surface distances, was recently proposed in \cite{emig}. It is based on a systematic {\it derivative expansion} of the PS potential, that generalizes an analogous expansion which was successfully used ~\cite{fosco2,bimonte3,bimonte4} to study the Casimir interaction between two non-planar  surfaces.   It has also been applied to other problems involving short range interactions between surfaces, like radiative heat transfer~\cite{golyk} and stray electrostatic forces between conductors~\cite{fosco3}.  From this expansion it was possible to obtain the leading and the next-to-leading curvature corrections to the PS interaction.

In current CP experiments, it is often  of interest to understand how the atom-surface interaction is modified by the temperature of the surface and/or of the environment. The temperature dependence of the CP interaction has been first demonstrated in the experiment \cite{obrecht} involving a magnetically trapped $^{87}{\rm Rb}$ Bose-Einstein condensate placed at a distance of a few microns from a heated fused silica substrate. Another very recent experiment \cite{laliotis}  probed the thermal CP potential between ${\rm Cs}^*(7 {\rm D}_{3/2})$ atoms and a hot sapphire substrate at a distance around 100 nm, in the temperature range from 500 to 1,000 K, paving the way to the control of atom-surface interactions by thermal fields. 
Temperature corrections to the PS potential for {\it planar} surfaces  have been studied before by several authors \cite{wu,mauro1,mauro2,harber,mendes,buh1,bezerra,galina2,lev,dere,zhu1,ellin1,ellin2,ellin3,zhu2},  both at equilibrium and out of equilibrium. For non-planar surfaces, the only study known to us is \cite{ellin4}, where  the thermal PS potential for a particle near a microsphere
has been studied using a multipole expansion.  In this paper we use the derivative expansion developed in \cite{emig} to study the interplay of curvature and temperature in the PS potential of an anisotropic small particle in thermal equilibrium with a gently curved surface of any shape.  For simplicity we consider here the idealized case of a perfectly conducting surface, and we postpone the study of  dielectric surfaces to a future work.  
 
The paper is organized as follows. In Sec.~II we derive the general structure of the derivative expansion at finite temperature for an arbitrary material surface, and simplify the results for a perfectly conducting surface. In Sec.~III  curvature corrections for the retarded Casimir-Polder interaction are computed for an anisotropic particle in a small temperature expansion and in the classical high temperature limit. Implications and possible extensions of our results are discussed in Sec.~IV.

\section{Derivative expansion of the particle-surface potential}

\begin{figure} 
\includegraphics [width=.9\columnwidth]{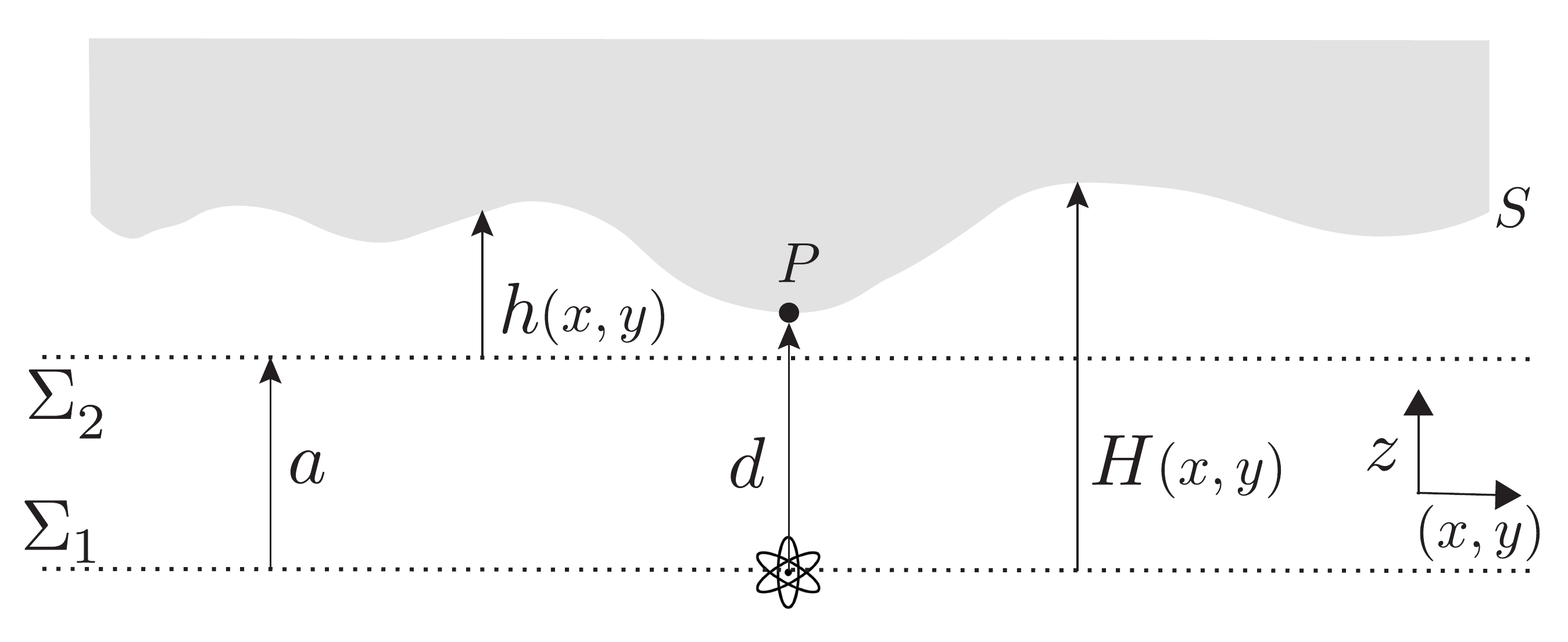} 
\caption{\label{fig1} Coordinates parametrizing a configuration  consisting of  a small particle near a gently curved   surface.}  
\end{figure}

We consider a particle (an `atom,'  a molecule, or any polarizable micro-particle) that is in thermal equilibrium with a dielectric surface $S$ at temperature $T$. We assume that the particle is small enough (compared to the scale of its
separation $d$ to the surface), such that its response to em fields is fully described by the dynamic electric dipolar polarizability tensor $\alpha_{\mu \nu}(\omega)$. (We assume for simplicity that the particle has a negligible magnetic polarizability, as is usually the case). Let $\Sigma_1$ be the plane through the particle which is orthogonal to the distance vector (which we take to be the ${\hat {\bf z}}$ axis) connecting the particle to the point $P$ of $S$ closest to the particle.  We assume that the surface $S$ is characterized by a {\it smooth} profile $z=H({\bf x})$, where ${\bf x} =(x,y)$ is the vector spanning $\Sigma_1$, with origin at the particle's position (see Fig.~\ref{fig1}). In what follows greek indices $\mu, \nu, \dots$ label all coordinates $(x,y,z)$, while latin indices $i,j,k \dots$ refer to $(x,y)$ coordinates in the plane $\Sigma_1$. Throughout we adopt the convention that repeated indices are summed over.  

The exact interaction potential at finite temperature $T$  is given by the scattering  formula~\cite{sca1,sca2}
\begin{equation}
\label{eq.4}
U= - k_B T \sideset{}{'}\sum_{n=0}^\infty \text{Tr} \, [ {\mathbb T}^{(S)} {\mathbb U} {\mathbb T}^{(A)}  {\mathbb U}  ](\kappa_n) \, .
\end{equation} 
Here ${\mathbb T}^{(S)}$ and ${\mathbb T}^{(A)}$ denote, respectively, the T-operators of the plate $S$ and the particle, evaluated at the Matsubara wave numbers $\kappa_n=2\pi n k_BT/(\hbar c)$, and the primed sum indicates that the $n=0$ term carries weight $1/2$.  In a plane-wave basis $|{\bf k},Q \rangle$~\cite{fn3} where ${\bf k}$ is the in-plane wave-vector, and $Q=E,M$ denotes respectively electric (transverse magnetic) and magnetic (transverse electric) modes, the translation operator $\mathbb U$ in Eq.~(\ref{eq.4}) is diagonal with matrix elements $e^{-d q}$ where $q=\sqrt{{ k}^2+ \kappa_n^2}\equiv q(k)$, $k=|{\bf k}|$. The matrix elements of the T-operator for the small particle in the dipole approximation are \be { {\cal T}}_{QQ'}^{(A)}({\bf k}, {\bf k}' ) =-\frac{2 \pi \kappa_n^2}{ \sqrt{q q'}} e^{(+)}_{Q \mu}({\bf k}) \alpha_{\mu \nu}({\rm i} c\kappa_n) e^{(-)}_{Q' \nu}({\bf k}')\;, \ee where $q'=q(k')$, ${\bf e}^{(\pm)}_{M}({\bf k})={\hat {\bf z}} \times {\hat {\bf k} }$ and ${\bf e}^{(\pm)}_{E}({\bf k})=-1/\kappa_n ({\rm i} k {\hat {\bf z}} \pm q {\hat {\bf k} })$, with ${\hat {\bf k} }={\bf k}/k$.  There are no analytical formulae for the elements of the T-operator of a curved plate ${\mathbb T}^{(S)} $, and its computation is in general quite challenging, even numerically. However, it has been shown in Ref.~\cite{emig} that at $T=0$ and in the classical limit for any smooth surface it is possible to compute the leading curvature corrections to the potential in the experimentally relevant limit of small separations.  The key idea is that the PS interaction falls off rapidly with separation, and it is thus reasonable to expect that the potential $U$ is mainly determined by the geometry of the surface $S$ in a small neighborhood of the point $P$ of $S$ which is {\it closest} to the particle. This physically plausible idea suggests that for small separations $d$ the potential $U$ can be expanded as a series expansion in an increasing number of derivatives of the height profile $H$, evaluated at the particles's position. Here we extend this idea to compute thermal corrections to the potential at small separations.
Up to fourth order, and assuming that the surface is homogeneous and isotropic, the most general expression which is invariant under rotations of the $(x,y)$ coordinates, and that involves at most four derivatives of $H$ (but no first derivatives since $\nabla H ({\bf 0})=0$) can be expressed (up to ${\cal O}(d^{-1}$)) in the form 
\begin{widetext}
\begin{align}
\label{derexpa}
U & = -\frac{k_B T}{d^3} \sideset{}{'}\sum_{n=0}^{\infty} \left\{\beta^{(0)}_1 \alpha_\perp + \beta^{(0)}_2 \alpha_{zz} + d\times\left[ (\beta^{(2)}_1 \alpha_\perp + \beta^{(2)}_2 \alpha_{zz}) \nabla^2 H+
 \beta^{(2)}_3  \left(\partial_i \partial_j H - \frac{1}{2} \nabla^2 H \delta_{ij}\right) \alpha_{ij}\right]+d^2\times \right. \\ \nonumber
& \left[\beta^{(3)}  \alpha_{zi} \partial_i \nabla^2 H \!
 \left.+(\nabla^2 H)^2 (\beta^{(4)}_1 \alpha_\perp\! + \beta^{(4)}_2 \alpha_{zz} )+  (\partial_i \partial_j H)^2 (\beta^{(4)}_3 \alpha_\perp\! + \beta^{(4)}_4 \alpha_{zz} )+ \beta^{(4)}_5
\nabla^2 H \! \left(\partial_i \partial_j H - \frac{1}{2} \nabla^2 H \delta_{ij}\right) \alpha_{ij} \right]\right\}\,,
\end{align}
\end{widetext}
where $\alpha_\perp=\alpha_{xx}+\alpha_{yy}$, and it is understood that all derivatives of $H({\bf x})$ are evaluated at the atom's position, i.e., for ${\bf x}={\bf 0}$. The Matsubara sum runs now over the rescaled wave numbers $\xi_n=\kappa_n d$ at which the dimensionless coefficient functions $\beta^{(p)}_p(\xi)$ are evaluated.
For a general dielectric surface these functions depend also on 
any other dimensionless ratio of frequencies characterizing the material of the surface. The derivative expansion  in Eq.~(\ref{derexpa}) can be formally obtained by a re-summation of the perturbative series for the potential for small in-plane momenta ${\bf k}$ \cite{emig}. We note that there are additional terms involving four derivatives of $H$ which, however, yield contributions $\sim 1/d$ (as do terms involving five derivatives of $H$) and are hence neglected.

The coefficients $\beta^{(p)}_q$ in Eq.~(\ref{derexpa}) can be extracted from  
the perturbative series of the potential $U$, carried to second order in the deformation $h({\bf x})$, which  in turn involves an expansion of the T-operator of the surface $S$ to the same order.
The latter expansion was worked out in Ref.~\cite{voron} for a dielectric surface characterized by a frequency dependent permittvity $\epsilon(\omega)$. It reads
\begin{align}
& { {\cal T}}_{QQ'}^{(S)}({\bf k}, {\bf k}' )=(2 \pi)^2 \delta^{(2)}({\bf k}-{\bf k'})\,\delta_{QQ'}\, r^{(S)}_{Q} (i c\kappa_n,{\bf k})
\nonumber\\
&+ \sqrt{q\,q'}\,\left[-2 \,B_{QQ'}({\bf k}, {\bf k}')\,\tilde{h}({{\bf k}- {\bf k}'})\right. \\
 & \left. + \!\!\int \!\!\frac{d^2 {\bf k}''}{(2 \pi)^2} (B_2)_{QQ'}({\bf k}, {\bf k}';{\bf k}'') \tilde{h}({{\bf k}\!- \!{\bf k}''}) \tilde{h}({{\bf k}''\!-\! {\bf k}'})+\dots \right]\;, \nonumber
\end{align}
where $r^{(S)}_{Q} (i c\kappa_n,{\bf k})$ denotes the familiar Fresnel reflection coeffcient of a flat surface.  Explicit expressions for $B_{Q Q'}({\bf k}, {\bf k}')$ and $(B_2)_{Q Q}({\bf k}', {\bf k}';{\bf k}'')$ are given in Ref.~\cite{voron}. The computation of the  coefficients $\beta^{(p)}_q$ involves an integral over ${\bf k}$ and ${\bf k'}$ (as it is apparent from Eq.~(\ref{eq.4})) that cannot be performed analytically for a dielectric plate, and has to be  estimated  numerically. In this paper, we shall content ourselves to considering the case of a perfect conductor, in which case the integrals can be performed analytically.  For a perfect conductor, the matrix $B_{QQ'}({\bf k}, {\bf k}')$ takes the simple form
\be
B({\bf k}, {\bf k}')=\left(\begin{array}{cc} \frac{{\hat{\bf k}}\cdot{\hat {\bf k}}'  \kappa_n^2+k k'}{ q q'}& \frac{\kappa_n}{ q} {\hat {\bf z}}\cdot({\hat{\bf k}}\times {\hat {\bf k}}') \\ \frac{\kappa_n}{ q'} {\hat {\bf z}}\cdot({\hat{\bf k}}\times {\hat {\bf k}}') & -{\hat{\bf k}}\cdot{\hat {\bf k}}' \\ \end{array} \right)\;,
\ee 
where the matrix indices $1,2$ correspond to $E,M$ respectively. 
 For perfect conductors, the matrix $(B_2)_{QQ'}({\bf k}, {\bf k}';{\bf k}'') $ is simply related to $B$ by
\be
(B_2)({\bf k}, {\bf k}';{\bf k}'') =2 q'' B({\bf k}, {\bf k}'') \sigma_3 B({\bf k}'', {\bf k}')\;,
\ee
where $\sigma_3={\rm diag}(1,-1)$. For perfect conductors the coefficients $\beta^{(p)}_q$ are functions of $\xi$ only, and we list them in  Table~\ref{tab:betas}.

\begin{widetext}
\begin{table*}
\begin{tabular}{|c|c|l|l|}
\hline
p & q & $\times e^{-2\xi}$  & $\times \text{Ei}(2\xi)$ \\
\hline
0 & 1 &  $\frac{1}{8}(1+2 { \xi}+4 { \xi}^2)$ &  $0$\\
 & 2 & $\frac{1}{4}(1+2 { \xi} )$ & $0$\\
2 & 1 & $-\frac{1}{32}(3+6 { \xi} +6 { \xi}^2+4 { \xi}^3 )$ & $-\frac{ { \xi}^4}{4} $\\
 & 2 & $-\frac{1}{16}(1+2 { \xi} -2 { \xi}^2+4 { \xi}^3 )$ & ${ \xi}^2 \left(1-\frac{{ \xi}^2}{2} \right)$\\
 & 3 & $-\frac{1}{32}(3+6 { \xi} +2{ \xi}^2-4 { \xi}^3 )$  & $\frac{ { \xi}^4}{4}$\\
3 & & $\frac{1}{32}(1+2 { \xi} -2{ \xi}^2+4 { \xi}^3 )$ & $-\frac{ { \xi}^2}{4} (2-{ \xi}^2)$\\
4 & 1 & $\frac{1}{384}(3+6 { \xi} +15{ \xi}^2+22 { \xi}^3+2 { \xi}^4-4{ \xi}^5 )$ & $\frac{ { \xi}^4}{48} (6-{ \xi}^2)$\\
 & 2 &  $-\frac{1}{960}(15+542 { \xi} +259{ \xi}^2-546 { \xi}^3-14 { \xi}^4+28{ \xi}^5 )$ & $-\xi^2(2 - \frac{7\xi^2}{6} + \frac{7\xi^4}{120})$\\
 & 3 &  $\frac{1}{192}(15+30 { \xi} -9{ \xi}^2+70 { \xi}^3+2 { \xi}^4-4{ \xi}^5 )$ & $\frac{ { \xi}^4}{24} (18-{ \xi}^2) $\\
 & 4 &  $\frac{1}{480}(45+218 { \xi} -59{ \xi}^2+146 { \xi}^3+14 { \xi}^4-28{ \xi}^5 )$ & $\frac{ { \xi}^4}{60} (40-7{ \xi}^2)$\\
 & 5 &  $\frac{1}{96}(9+18 { \xi} -27{ \xi}^2+50 { \xi}^3-2 { \xi}^4+4{ \xi}^5 )$ & $ { \xi}^4\left(1+\frac{{ \xi}^2}{12}\right)$\\
\hline
\end{tabular}
\caption{\label{tab:betas} The coefficients $\beta^{(p)}_q(\xi)$ are obtained by multiplying the third column by $e^{-2\xi}$, and adding the fourth column times $\text{Ei}(2\xi)=-\int_{2\xi}^\infty dt \exp(-t)/t$.}
\end{table*}

The potential in Eq.~(\ref{derexpa}) can be expressed also in the radii of curvature, $R_1$ and $R_2$, of the surface $S$ at $P$.  The local expansion of $H$ can be chosen as 
$H=d+x^2/(2 R_1)+ y^2/(2 R_2)+\cdots$ so that the derivative expansion of $U$ reads
\begin{align}
U & =-\frac{k_B T}{d^3} \sideset{}{'}\sum_{n=0}^\infty\left\{\beta^{(0)}_1 \alpha_\perp + \beta^{(0)}_2 \alpha_{zz} + \left(\frac{d}{R_1}+\frac{d}{R_2} \right) (\beta^{(2)}_1 \alpha_\perp + \beta^{(2)}_2 \alpha_{zz})+
 \frac{\beta^{(2)}_3}{2}  \left(\frac{d}{R_1}-\frac{d}{R_2}\right) (\alpha_{xx}-\alpha_{yy}) \right.\nonumber \\
&+ d^2 \beta^{(3)}  \alpha_{zi} \partial_i \left(\frac{1}{R_1}
+\frac{1}{R_2} \right) 
+ \left(\frac{d}{R_1}+\frac{d}{R_2} \right)^2 (\beta^{(4)}_1 \alpha_\perp + \beta^{(4)}_2 \alpha_{zz} )\frac{}{}
\nonumber \\
& \left.\left.
+ \left[\left(\frac{d}{R_1}\right)^2+\left(\frac{d}{R_2} \right)^2 \right]   (\beta^{(4)}_3 \alpha_\perp + \beta^{(4)}_4 \alpha_{zz} )+ \frac{\beta^{(4)}_5}{2} \left[\left(\frac{d}{R_1}\right)^2-\left(\frac{d}{R_2} \right)^2  \right] (\alpha_{xx}-\alpha_{yy}) 
  \right\}\right.\; ,\label{derexpa2}
\end{align}
\end{widetext}
where the dependence of the coefficients $\beta^{(p)}_q(\xi)$ on $\xi_n$ is again not shown explicitly.

\section{Retarded Particle-Surface potential}

The $\beta$ coefficients in Eq.~(\ref{derexpa}) are significantly different from zero only
for rescaled  frequencies $\xi_n \lesssim 1$.  Therefore, for separations small compared to the radii of surface curvature  but $d \gg c/\omega_r$, where $\omega_r$ is a
typical resonance frequency of the particle (atom), we can replace $\alpha_{\mu \nu}({\rm i} c\kappa)$ in Eqs.~(\ref{derexpa},\ref{derexpa2}) by its static limit $\alpha_{\mu \nu}(0) \equiv \alpha_{\mu \nu}^0$. 
Then a small temperature expansion of the potential can be obtained by evaluating the sum using the Abel-Plana formula. This yields for $d \ll \lambda_T=\hbar c/(2 \pi k_B T)$ the {\it retarded} Casimir-Polder potential
\be
U_{\rm CP}=-\frac{\hbar c}{ \pi d^4}\left[\alpha_{\perp}^0 \eta_{\perp}+\alpha_{zz}^0 \eta_{zz}+\alpha_{zi}^0 \eta_{zi}+(\alpha_{xx}^0-\alpha_{yy}^0) \eta_{xy} \right] \, .
\ee
We performed the temperature expansion up to order $(d/\lambda_T)^5$. This order is the smallest one for which thermal corrections become visible in all terms of the following coefficients that describe curvature corrections up to order $(d/R_i)^2$,
\begin{widetext}
\begin{align}
\label{eq:eta1}
  \eta_{\perp}&=\frac{1}{8}-\frac{1}{360} \left(\frac{d}{\lambda_T}\right)^4-\left(\frac{d}{R_1}+\frac{d}{R_2} \right)\left[\frac{3 }{40}-\frac{3 \zeta(5)}{32 \pi^4} \left(\frac{d}{\lambda_T}\right)^5\right]
  +\left(\frac{d}{R_1}+\frac{d}{R_2} \right)^2  \left[\frac{3 }{280}- \frac{3 \zeta(5)}{64 \pi^4} \left(\frac{d}{\lambda_T}\right)^5\right]  \nonumber \\\
  &+\left[\left(\frac{d}{R_1}\right)^2\!+\left(\frac{d}{R_2} \right)^2 \right]   \left[\frac{13 }{280}+\frac{1}{360} \left(\frac{d}{\lambda_T}\right)^4\right] \\\
\label{eq:eta2}
 \eta_{zz}&= \frac{1}{8}+\frac{1}{360} \left(\frac{d}{\lambda_T}\right)^4 -\left(\frac{d}{R_1}+\frac{d}{R_2} \right) \left[
\frac{1}{15}-\frac{\zeta(3)}{8 \pi^2} \left(\frac{d}{\lambda_T}\right)^3+\frac{1}{90} \left(\frac{d}{\lambda_T}\right)^4- \frac{3\zeta(5)}{16 \pi^4} \left(\frac{d}{\lambda_T}\right)^5 \right] \nonumber \\\
&- \left(\frac{d}{R_1}+\frac{d}{R_2} \right)^2 \left[\frac{1}{240}-\frac{1}{45} \left(\frac{d}{\lambda_T}\right)^2+\frac{\zeta(3)}{4 \pi^2} \left(\frac{d}{\lambda_T}\right)^3-\frac{1}{60} \left(\frac{d}{\lambda_T}\right)^4+\frac{7 \zeta(5)}{16 \pi^4} \left(\frac{d}{\lambda_T}\right)^5\right] \nonumber \\\
&+\left[\left(\frac{d}{R_1}\right)^2\!+\left(\frac{d}{R_2} \right)^2 \right] \left[\frac{3 }{40}-\frac{1}{90} \left(\frac{d}{\lambda_T}\right)^2+\frac{1}{180} \left(\frac{d}{\lambda_T}\right)^4-\frac{\zeta(5)}{4 \pi^4} \left(\frac{d}{\lambda_T}\right)^5 \right]\\\
\label{eq:eta3}
\eta_{zi}&= d^2  \partial_i \left(\frac{1}{R_1}
+\frac{1}{R_2} \right) \left[\frac{1}{30} -\frac{ \zeta(3)}{16 \pi^2} \left(\frac{d}{\lambda_T}\right)^3+\frac{1}{180} \left(\frac{d}{\lambda_T}\right)^4-\frac{3 \zeta(5)}{32 \pi^4} \left(\frac{d}{\lambda_T}\right)^5\right] \\\
\label{eq:eta4}
\eta_{xy}&=-  \left(\frac{d}{R_1}-\frac{d}{R_2}\right)  \left[ \frac{1}{40}+\frac{3 \zeta(5)}{64 \pi^4} \left(\frac{d}{\lambda_T}\right)^5\right]+\left[\left(\frac{d}{R_1}\right)^2\!-\left(\frac{d}{R_2} \right)^2  \right]  \left[\frac{9}{560}+\frac{1}{360} \left(\frac{d}{\lambda_T}\right)^4-\frac{3 \zeta(5)}{16 \pi^4} \left(\frac{d}{\lambda_T}\right)^5\right] \, .
\end{align}
\end{widetext}
In the special case of a spherical atom near a cylindrical metallic shell at $T=0$, the above formula is in full agreement with  Eq.~(30) of Ref.~\cite{galina},  as well as with Eqs.~(25)-(26) and Eqs.~(38)-(39) or Ref. \cite{pablo}. 
The magnitude of the thermal corrections depends on the component of the polarizability tensor and the order of the curvature correction. The thermal contributions proportional to $\alpha^0_{zz}$ scales as $T^3$ at order $1/R_i$ and as $T^2$ at order $1/R_i^2$.  The thermal terms proportional to lateral $(x,y)$ components of $\alpha^0$ are each reduced by $T^2$ compared those of $\alpha^0_{zz}$.
This constitutes a manifestation of correlations between curvature and thermal effects in fluctuation induced interactions.

For completeness, we consider also the classical high temperature limit,
where the  Casimir free energy is given by the first term of the Matsubara sum in Eq.~\eqref{eq.4} only.
From the limit $\kappa\to 0$ of the coefficients $\beta^{(p)}_q$ we obtain the classical free energy to order $(d/R_i)^2$ as
\begin{widetext}
\begin{align}
\label{eq:classical}
U_{\rm classical}& = -\frac{k_B T}{2} \frac{1}{d^3} \left\{   \frac{1}{8} \alpha^0_\perp + \frac{1}{4} \alpha^0_{zz} 
-\frac{3}{64} \left( 3\frac{d}{R_1} + \frac{d}{R_2}\right) \alpha^0_{xx} 
-\frac{3}{64} \left( \frac{d}{R_1} + 3\frac{d}{R_2}\right) \alpha^0_{yy} 
-\frac{1}{16} \left(  \frac{d}{R_1}+\frac{d}{R_2}\right) \alpha^0_{zz}   \right. \\
&\!\!\!\!\!\!\! \left. +\frac{1}{128} \left( 17 \frac{d^2}{R_1^2} + 5 \frac{d^2}{R_2^2} +2 \frac{d^2}{R_1R_2}\right) \alpha^0_{xx}   
+\frac{1}{128} \left( 17 \frac{d^2}{R_2^2} + 5 \frac{d^2}{R_1^2} +2 \frac{d^2}{R_1R_2}\right) \alpha^0_{yy}
+\frac{1}{64} \left( 5\frac{d^2}{R_1^2} +5\frac{d^2}{R_2^2} -2\frac{d^2}{R_1R_2}\right) \alpha^0_{zz} \right\} \, . \nonumber
\end{align}
\end{widetext}
The corrections to this limit due to higher order Matsubara terms decay exponentially in $d/\lambda_T \sim T$.

\begin{figure}[h]
\includegraphics [width=.8\columnwidth]{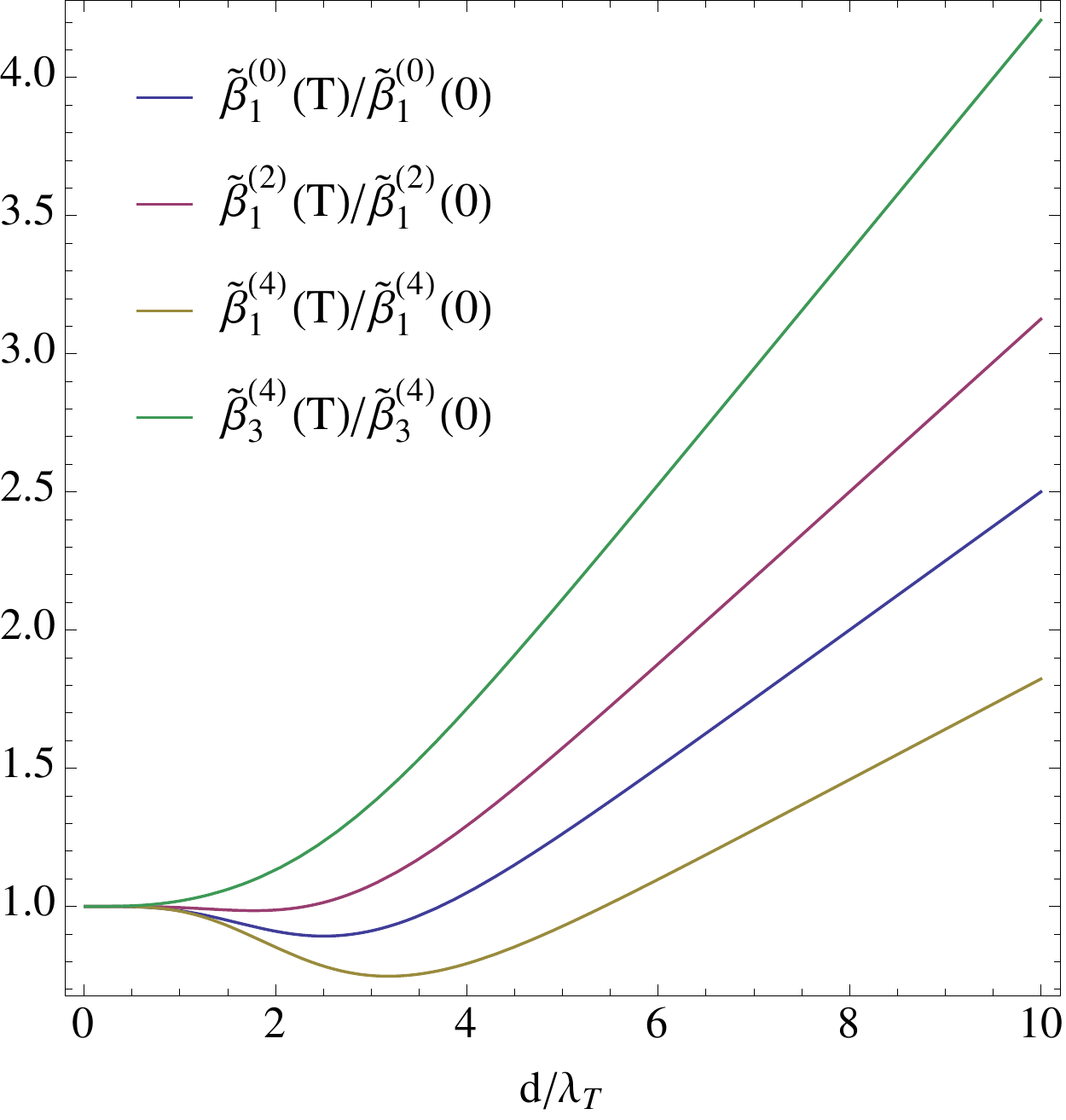} 
\caption{\label{fig2} Temperature dependence of the coefficients $\tilde\beta^{(p)}_q(T)=(d/\lambda_T) 
\sideset{}{'}\sum_{n=0}^\infty \beta^{(p)}_q(n d/\lambda_T)$ that mutliply $\alpha^0_\perp$ in Eq.~\eqref{derexpa2}, normalized to $\tilde\beta^{(p)}_q(T=0)=\int_0^\infty \beta^{(p)}_q(\xi) d\xi$.}  
\end{figure}
\begin{figure}[h]
\includegraphics [width=.8\columnwidth]{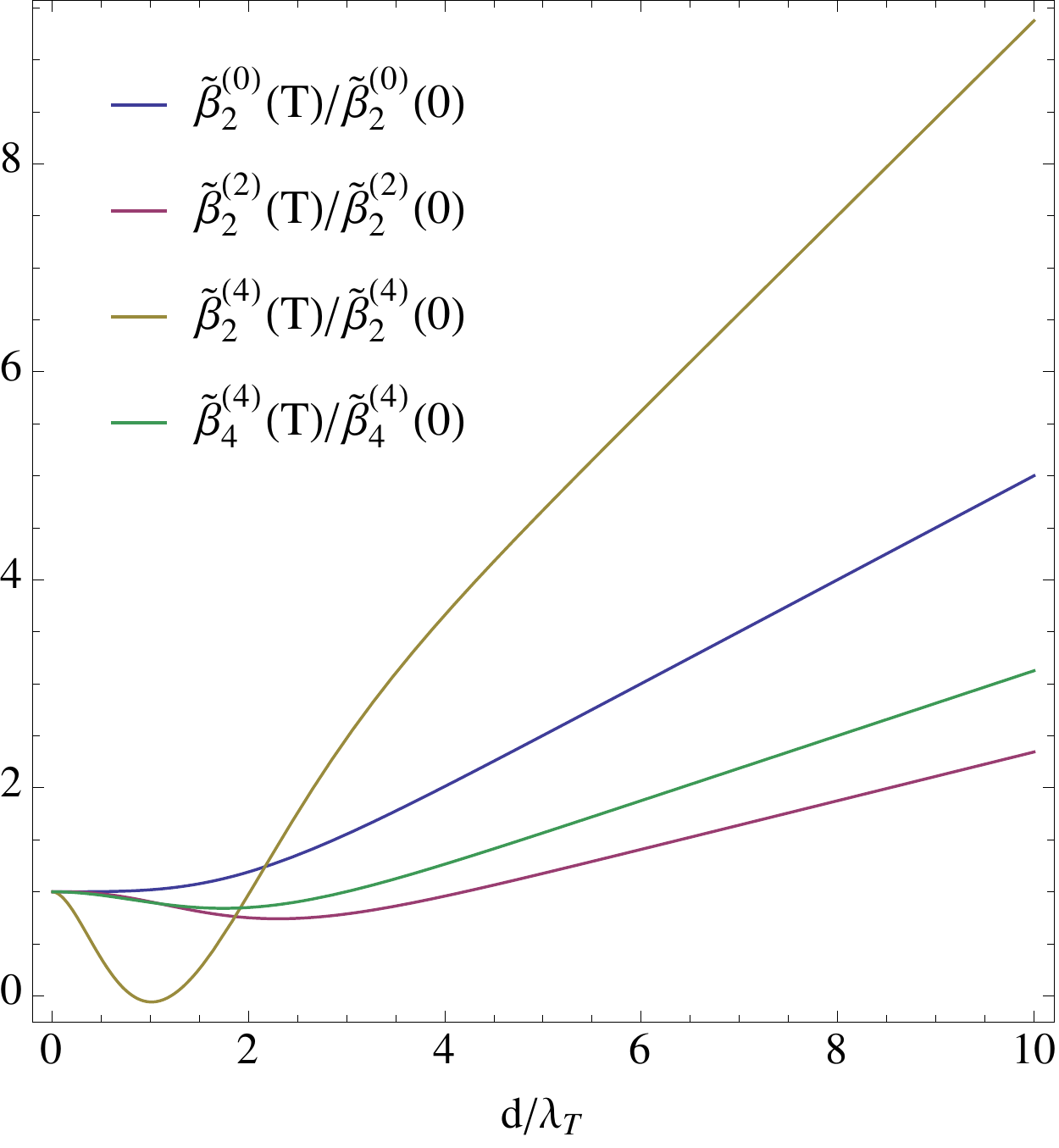} 
\caption{\label{fig3} Same as Fig.~\ref{fig2} but for the coefficients of $\alpha^0_{zz}$ in Eq.~\eqref{derexpa2}.}  
\end{figure}
\begin{figure}[h]
\includegraphics [width=.8\columnwidth]{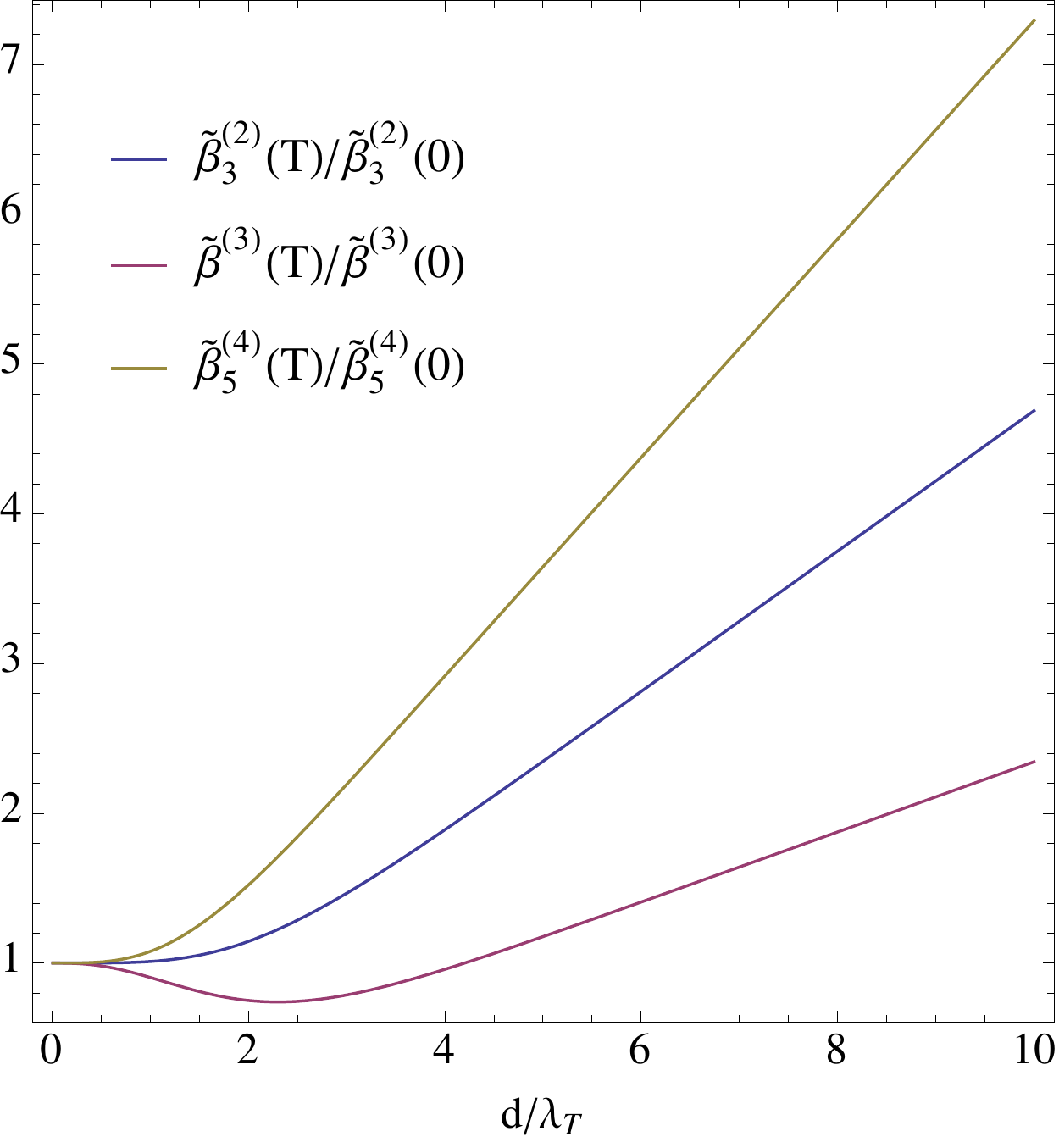} 
\caption{\label{fig4}Same as Fig.~\ref{fig2} but for the coefficients of the mixed components of $\alpha^0$ in Eq.~\eqref{derexpa2}.}  
\end{figure}

It is instructive to study the {\it full} temperature dependence of the curvature corrections for $U$ 
and how it compares to $T$-dependence of the curvature independent terms. To this end we compute numerically
the Matsubara sums $\tilde\beta^{(p)}_q(T)=(d/\lambda_T) 
\sideset{}{'}\sum_{n=0}^\infty \beta^{(p)}_q(n d/\lambda_T)$ and plot them normalized to their zero temperature limit 
$\tilde\beta^{(p)}_q(T=0)=\int_0^\infty \beta^{(p)}_q(\xi) d\xi$ as function of $d/\lambda_T$ in Figs.~\ref{fig2}-\ref{fig4}.
These sums appear in Eq.~\eqref{derexpa2} in the retarded limit where the tensor $\alpha$ is frequency independent.
At large temperature all coefficients show a linear temperature dependence, reproducing the classical result of Eq.~\eqref{eq:classical}. At small $T$ the coefficient display only a weak temperature dependence which is consistent with the powers of $T$ in Eqs.~\eqref{eq:eta1}-\eqref{eq:eta4}. An exceptionally strong temperature dependence is observed for the coefficient $\beta^{(4)}_2$, see Fig.~\ref{fig3}.

\section{Summary and implications}

We have studied the combined effect of thermal fluctuations and surface curvature on the Casimir-Polder interaction at small particle-surface separations. Explicit results for the interaction potential have been obtained in the retarded limit 
for a perfectly conducting surface. We have presented analytical results at low and high temperatures, and numerical results at arbitrary temperatures. The employed gradient expansion allows for interesting extensions to dielectric surfaces, and more sophisticated models for the particle, including excited atoms or molecules. 

Our work opens the prospect of studying a number a novel phenomena in particle-surface interactions, arising from the interplay of thermal excitations and surface shape.  It is easy to deduce already from Eq.~(\ref{derexpa2}) that curvature of the surface can exert a torque, rotating an anisotropic particle into a specific low energy orientation.  The distinct temperature dependencies of the different polarisation components shown in Figs.~\ref{fig2}-\ref{fig4} suggest that the preferred orientation at a fixed distance may change upon heating or cooling the system. Non-equilibrium situations, involving an excited (Rydberg) atom or a polar molecule, or a surface held at a different temperature also provide additional avenues for exploration. Recent experiments \cite{laliotis} have demonstrated for planar surfaces a substantial increase of the interaction due to thermally excited surface waves that couple to atomic transitions. The sensitivity of surface-polariton modes to surface curvature suggests the possibility to tailor thermal fluctuations so that they coincide with transitions of the particle, leading to an increased interaction. This idea could be realised by the use of designed nano-structured surfaces.

\begin{acknowledgments}
We thank M.~Kardar for valuable discussions.  
\end{acknowledgments}

\end{document}